\documentclass[11pt]{article}
\usepackage{amsmath,amssymb}
\usepackage{hyperref}
\newcommand{\bea}{\begin{eqnarray}}
\newcommand{\eea}{\end{eqnarray}}

\newcommand{\ee}{\mathrm{e}}
\newcommand{\im}{\mathrm{i}}

\renewcommand{\Im}{\operatorname{Im}}

\newsavebox{\uuunit}
\sbox{\uuunit}
    {\setlength{\unitlength}{0.825em}
     \begin{picture}(0.6,0.7)
        \thinlines
        \put(0,0){\line(1,0){0.5}}
        \put(0.15,0){\line(0,1){0.7}}
        \put(0.35,0){\line(0,1){0.8}}
       \multiput(0.3,0.8)(-0.04,-0.02){10}{\rule{0.5pt}{0.5pt}}
     \end {picture}}

\numberwithin{equation}{section}
\begin{document}
\begin{titlepage}
\begin{center}
\hfill LMU-ASC 01/07 \\
\hfill MPP-2007-7 \\
\hfill {\tt hep-th/0701176}\\
\vskip 6mm

%%%%%%%%%%%%%%%%%%%%%%%%%%%%%%%%%%%%%%%%%%
{\Large \textbf{Entropy function for rotating extremal black holes \\
\vskip 1mm
in very special geometry}}
%%%%%%%%%%%%%%%%%%%%%%%%%%%%%%%%%%%%%%%%%%%
\vskip 8mm

\textbf{G. L.~Cardoso$^{a}$, J. M. Oberreuter$^a$ and J.~Perz$^{a,b}$}

\vskip 4mm
$^a${\em Arnold Sommerfeld Center for Theoretical Physics\\
Department f\"ur Physik,
Ludwig-Maximilians-Universit\"at M\"unchen \\ Theresienstr. 37, 
80333 M\"unchen, Germany}
\\[1mm]
$^b${\em Max-Planck-Institut f\"ur Physik\\
F\"ohringer Ring 6, 80805 M\"unchen, Germany
}\\[1mm]

\centerline{
{\tt gcardoso,oberreuter,perz@theorie.physik.uni-muenchen.de}}

\end{center}
\vskip .2in
%%%%%%%%%%%%%%%%%%%%%%%%%%%%%%%%%%%%%%%%%%%%%%%%%%%%%
\begin{center} {\bf ABSTRACT } \end{center}
\begin{quotation}\noindent
We use the relation between extremal black hole solutions in five- and in
four-dimensional $N=2$ supergravity theories with cubic prepotentials
to define the entropy function for extremal black holes 
with one angular momentum in five dimensions.
We construct two types of solutions to the associated attractor equations.

\end{quotation}

\vfill
%%%%%%%%%%%%%%%%%%
%\flushleft{\today}
%%%%%%%%%%%%%%%%%%
\end{titlepage}
%%%%%%%%%%%%%%%%
\eject
%%%%%%%%%%%%%%%%
%%%%%%%%%%%%%%%%%%%%%%%%%%%%%%%%%%%%%%%%%%%%%%%%%%%%%%%%%%%%%%%%%%%%%%
%%%%%%%%%%%%%%%%%%%%%%%%%%%%%%%%%%%%%%%%%%%%%%%%%%%%%%%%%%%%%%%%%%%%%%
\section{Introduction}
\setcounter{equation}{0}
%%%%%%%%%%%%%%%%%%%%%%%%%%%%%%%%%%%%%%%%%%%%%%%%%%%%%%%%%%%%%%%%%%%%%%

An important feature of extremal black holes in the presence of scalar
fields 
is that these fields
attain fixed values at the horizon 
which are determined by the black hole charges.
These values are found by solving a set of so-called attractor equations,
which were first given in 
\cite{Ferrara:1995ih,Strominger:1996kf,Ferrara:1996dd,Ferrara:1996um}
in the context of supersymmetric black holes.
The attractor equations
can be obtained from a variational principle based on 
an entropy function
\cite{Sen:2005wa,Astefanesei:2006dd}.  The value of the 
entropy function at the stationary point yields
the macroscopic entropy of the black hole.

Extremal black holes in five dimensions can be related to 
extremal black holes
in four dimensions.  This connection is implemented by placing the 
five-dimensional black hole in a Taub-NUT geometry, and by using the
modulus of the Taub-NUT space to interpolate between the five and the
four-dimensional description. 
In the vicinity of the NUT charge,
spacetime looks five-dimensional, whereas far away from the NUT the
spacetime looks four-dimensional.  This connection was first 
established in \cite{Gaiotto:2005gf,Gaiotto:2005xt}
for supersymmetric 
black holes in the context of $N=2$ supergravity theories that in
four dimensions are based on cubic prepotentials, 
and was further discussed in
\cite{Behrndt:2005he}.

In the following, we focus on rotating extremal 
black holes in five dimensions which
are connected to static extremal black holes in four dimensions in the way
described above.
We use this link to define the entropy function
for these rotating black hole solutions in the
context of 
$N=2$ supergravity theories with cubic prepotentials.  
In four dimensions, the static extremal black holes 
%in $N=2$ supergravity theories
we consider
carry charges $(P^I, Q_I)$, where  
$P^0 \neq 0$ corresponds to the NUT charge in five dimensions.
These four-dimensional 
black holes are connected to 
rotating five-dimensional black holes with 
one independent angular momentum parameter.
The five-dimensional $N=2$ supergravity
theories contain Chern-Simons terms for the abelian gauge fields, 
so that the definition of the entropy function given in 
\cite{Sen:2005wa,Astefanesei:2006dd} 
cannot be directly
applied whenever these terms play a role 
for the given background.  Therefore, we 
define the entropy function for these rotating
five-dimensional black holes to equal the entropy function of the associated
static black holes in four dimensions.  The latter was computed for 
$N=2$ supergravity theories in 
\cite{Sahoo:2006rp,Cardoso:2006xz}.
Then, we specialize to the case of 
black holes with non-vanishing charges $(P^0,Q_I)$, which in five
dimensions correspond to rotating electrically charged extremal 
black holes in a Taub-NUT geometry.  Extremization of the entropy function
yields a set of attractor equations for the various parameters characterizing
the near-horizon solution.  We check that these attractor equations are 
equivalent to
the equations of motion
in five dimensions evaluated in the black hole background.
We construct two types of solutions to the attractor equations and
we compute
their entropy.

Our approach for defining the entropy function in the
presence of Chern-Simons terms is based on dimensional reduction, and is 
therefore similar to the approach used in 
\cite{Sahoo:2006vz} for defining the entropy function of the three-dimensional
BTZ black hole.  Related results for rotating $AdS_5$ black holes have 
appeared in \cite{Morales:2006gm}.

\section{Extremal black holes in five and four dimensions \label{rel4and5} }

Extremal black holes in five dimensions can be connected to 
extremal black holes
in four dimensions, as described in the introduction.  
In the following, we focus on rotating black holes in five dimensions which
are connected to static black holes in four dimensions.
The associated
near-horizon geometries are related by dimensional reduction over a compact
direction of radius $R$.  In the context of 
five-dimensional theories based on $n$ abelian gauge fields
$A_5^A$ and real scalar fields $X^A$ ($A= 1, \dots, n$) coupled to gravity, 
the reduction is based on the following standard
formulae (see for instance \cite{Gunaydin:1983bi}),
\begin{eqnarray}
ds^2_5 &=& \ee^{2 \phi} \, ds^2_4 + \ee^{- 4 \phi} \,(d x^5 - A^0_4)^2 
\;\;\;,\;\; dx^5 = R \, d \psi \;,
\nonumber\\
A_5^A &=& A_4^A + C^A \, (d x^5 - A_4^0) \;,\nonumber\\
{\hat X}^A &=& \ee^{- 2 \phi} \, X^A\;,
\label{dic}
\end{eqnarray}
where the $A_4^I$ denote the four-dimensional abelian
gauge fields (with $I = 0, A$).

We will focus on $N=2$ supergravity theories that are based 
on cubic prepotentials in four dimensions.
As we review in appendix \ref{appred}, 
the rescaled scalar fields ${\hat X}^A$ and 
the Kaluza-Klein scalars $C^A$
are combined into the four-dimensional complex scalar fields $z^A$
\cite{Gunaydin:1983bi},
\begin{eqnarray}
z^A = C^A + \im {\hat X}^A \;.
\label{zA}
\end{eqnarray}
We take the fields $C^A$ and 
${\hat X}^A$, and hence also $z^A$,  to be dimensionless.

The near-horizon geometry of the rotating five-dimensional black hole
is taken to be a squashed $AdS_2 \times S^3$ given by \cite{Morales:2006gm}
\begin{eqnarray}
ds^2_5 = v_1 (- r^2 dt^2  + \frac{dr^2}{r^2}) 
+ \frac{v_2}{4}  \left(d \theta^2 + \sin^2 \theta \, d\varphi^2 \right)
+ \frac{v_2 v_3}{4} \left(d \psi + \cos \theta \, d \varphi
 - \alpha \, r dt\right)^2  \;,
\label{5dline2}
\end{eqnarray}
where $\theta \in [0,  \pi],\, \varphi \in [0, 2 \pi), \,
\psi \in [0, 4 \pi) $. The parameters $v_1, v_2, v_3$ and $\alpha$ are
constant.
The near-horizon geometry of the associated static four-dimensional
black hole is of the $AdS_2 \times S^2$ type,
\begin{eqnarray}
ds^2_4 = \tilde{v}_1 (- r^2 dt^2  + \frac{dr^2}{r^2}) 
+ \tilde{v}_2  \left(d \theta^2 + \sin^2 \theta \, d\varphi^2 \right)
\label{line4ext}
\end{eqnarray}
with constant parameters ${\tilde v}_1$ and ${\tilde v}_2$.
Using \eqref{dic}, we find the following relations,
\begin{eqnarray}
\ee^{- 4 \phi} &=& \frac{v_2 v_3}{4 R^2 } \;, \nonumber\\
A_4^0 &=& R \left( - \cos \theta d \varphi + \alpha r dt \right) \;,
\label{vpvv}
\end{eqnarray}
as well as 
\begin{eqnarray}
{\tilde v}_1 = v_1 \sqrt{\frac{v_2 v_3}{4 R^2}} \;\;\;,\;\;\;
{\tilde v}_2 = \frac{v_2}{4}  \sqrt{\frac{v_2 v_3}{4 R^2}} 
 \;,
\label{vs45}
\end{eqnarray}
and hence
\begin{eqnarray}
{\tilde v}_1 \, {\tilde v}_2 = \frac{v_1 \, v^2_2 \, v_3}{16 R^2} \;\;\;,\;\;\;
\frac{{\tilde v}_1}{{\tilde v}_2} =
4 \,\frac{v_1}{v_2}  \;.
\label{v1v2}
\end{eqnarray}
We denote the electric fields in four and five dimensions by
$F_{rt} = e$.
Hence, we rewrite $A_4^0$ as
\begin{eqnarray}
A^0_4 = e^0_4\, r\, dt - p^0 R \cos \theta d \varphi \;,
\label{a04}
\end{eqnarray}
with $e^0_4 = \alpha R $ and the NUT charge $p^0 = 1$.

\section{Entropy function for rotating extremal black holes
in five dimensions}

The entropy function of \cite{Sen:2005wa,Astefanesei:2006dd} is derived from
the reduced Lagrangian.  The reduced Lagrangian ${\cal F}$ is obtained
by evaluating the Lagrangian in the near-horizon black hole background
and integrating over the horizon.  In five and four dimensions,
\begin{eqnarray}
{\cal F}_5 &=& \int d \psi \, d\theta \, d \phi \, \sqrt{-G} \,
{\cal L}_5 \;, \nonumber\\
{\cal F}_4 &=&  \int d\theta \, d \phi \, \sqrt{-g} \, {\cal L}_4 \;.
\label{redact45}
\end{eqnarray}
In the presence of Chern-Simons terms, however, the definition of
the entropy function given in \cite{Sen:2005wa,Astefanesei:2006dd}
is not directly applicable
whenever these terms play a role 
for the given background.  This is the situation encountered in $N=2$
supergravity theories in five dimensions, but not in four dimensions.
Therefore, we use dimensional reduction to define the entropy function
${\cal E}_5$ 
for rotating black holes in five dimensions in terms of the entropy function
${\cal E}_4$ 
for the associated four-dimensional black holes,
\begin{equation}
{\cal E}_5 = {\cal E}_4 \;.
\label{e5=e4}
\end{equation}

\subsection{Rotating electrically charged black holes in five dimensions}

Here
we consider rotating electrically charged extremal black hole solutions
in $N=2$ supergravity theories in five dimensions.  The bosonic
part of the five-dimensional Lagrangian is given by \eqref{lag5}.
The black hole solutions carry NUT charge $p^0 = 1$ as well as electric
charges $q_A$.  The near-horizon solution is specified in terms of 
constant scalars $X^A$, 
the
line element \eqref{5dline2} and 
the five-dimensional gauge potentials $A^A_5$,
\begin{eqnarray}
A^A_5 = e^A_5 
\, r \, dt + C^A \, R \, \left( d \psi + \cos \theta d \varphi \right)
\;,
\label{gaugepot5}
\end{eqnarray}
where $F^A_{rt} = e^A_5$ denotes the electric field in five dimensions.
Both $e^A_5$ and $C^A$ are constant.

These five-dimensional rotating extremal black holes are connected to static
electrically charged extremal black holes in four dimensions with 
constant scalars $z^A$, 
line
element \eqref{line4ext} and four-dimensional gauge potentials $A^I_4$
given by \eqref{a04} and 
\begin{eqnarray}
A^A_4 = e^A_4 \, r \, dt \;.
\label{gaugepot4}
\end{eqnarray}
The five- and four-dimensional electric fields are related by
\begin{eqnarray}
e^A_5 = e^A_4 - C^A \, e^0_4 = e^A_4 - \alpha \, R \, C^A
\label{efs45}
\end{eqnarray}
according to \eqref{dic}.  In our conventions, the electric fields
in five and four dimensions have length dimension one.

As reviewed in appendix \ref{appred}, the five- and four-dimensional
actions \eqref{lag5} and \eqref{lag4} are identical upon dimensional
reduction over $x^5$, up to boundary terms which are usually discarded and 
which arise when
integrating the Chern-Simons term in \eqref{lag5} by parts.
However, when evaluating these actions in a background with 
constant $C^A$, as is the case
for the near-horizon solutions under consideration, they are not
any longer equal to one another.
Namely, 
evaluating the Chern-Simons term in \eqref{lag5}
for constant $C^A$, and using $F^A_5 = F^A_4 - C^A \, F^0_4$
(see \eqref{dic}), we obtain with the help of \eqref{coup45}
\begin{eqnarray}
C_{ABC} \, F^A_5 \wedge  F^B_5 \wedge A^C_5 &=& 
\frac12 \, R \, d \psi \, d^4x \, \sqrt{-g} \, 
{\rm Re} {\cal N}_{IJ} \, F^I_4 \, {\tilde F}^J_4
\\
&& - R \,C_{ABC} \,\left(C^A \, C^B \, 
F^C_4 - \frac23 \,C^A \,C^B \,C^C \,F^0_4 \right)
\wedge F^0_4 \wedge d \psi \;. \nonumber
\end{eqnarray}
Thus, the actions differ by
\begin{eqnarray}
8 \pi \left(S_5 - S_4 \right) &=& \frac{1}{6 G_4}
\int  d^4x \, \sqrt{-g} \, 
{\rm Re} {\cal N}_{IJ} \, F^I_4 \, {\tilde F}^J_4 \nonumber\\
&&+ \frac{1}{6 G_4} \int C_{ABC} \,\left(C^A \, C^B \, 
F^C_4 - \frac23 \,C^A \,C^B \,C^C \,F^0_4 \right)
\wedge F^0_4 \;.
\end{eqnarray}
Similarly, in the background specified by \eqref{gaugepot5}, the
reduced Lagrangians \eqref{redact45} differ by
\begin{eqnarray}
{\cal F}_5 - {\cal F}_4  = \frac{1}{12 G_4} \, R \, 
C_{ABC} \, C^A \, C^B \, 
e^C_4 \;.
\label{diffredact}
\end{eqnarray}
This has to be taken into account when using \eqref{e5=e4} to define
the entropy function in five dimensions in terms of ${\cal E}_4$.
The entropy function of static extremal black holes in four dimensions
is the Legendre transform of the reduced Lagrangian ${\cal F}_4$ with
respect to the electric fields and 
reads \cite{Sen:2005wa}
\begin{equation}
 {\cal E}_4 = 2 \pi \left( 
- \frac12 \, e^I_4 \, Q_I \, G_4^{-1/2} - {\cal F}_4 \right)
\;,
\label{entfunc4}
\end{equation}
where we denote the four-dimensional electric charges by $Q_I$.
The normalizations are as in \cite{Sahoo:2006rp,Cardoso:2006xz},
with the additional
$G_4^{-1/2}$ to ensure that ${\cal E}_4$ is dimensionless.
Using \eqref{efs45}, \eqref{diffredact} and
\eqref{e5=e4}, we now express \eqref{entfunc4}
as
\begin{eqnarray}
{\cal E}_5 &=& 
2 \pi \left[ - \frac12 \, \alpha \, \left( J 
+ R \, C^A \left( q_A \, G_5^{-1/3} - 
\frac{2 \pi \, R^2}{3 G_5} \, C_{ABC} \, C^B \, C^C  \right) \right)
\right. \nonumber\\
&& \left. \qquad - \frac12 \, e^A_5 
\left( q_A \, G_5^{-1/3} - 
\frac{2 \pi \, R^2}{3 G_5} \, C_{ABC} \, C^B \, C^C  \right)
- {\cal F}_5 \right]
\;,
\label{entropyf5}
\end{eqnarray}
where the five-dimensional quantities $(J,q_A)$ are given in terms of the
four-dimensional electric charges $(Q_0, Q_A)$ by
\begin{eqnarray}
J &=& Q_0 \, R \, G_4^{-1/2} \;,
\nonumber\\
q_A \, G_5^{-1/3} &=& Q_A \, G_4^{-1/2}\;.
\label{char45}
\end{eqnarray}
In \eqref{valuealp} below, $J$ will be related to 
the angular momentum in five-dimensions.
Observe that in the presence of the $C^A$, the electric charges
$q_A$ are shifted by a term proportional to $C_{ABC} \, C^B \, C^C$.
This shift, which is due to \eqref{diffredact} and thus has its origin
in the presence of the Chern-Simons term in the five-dimensional action
\eqref{lag5}, has also been observed in \cite{Morales:2006gm}.  In addition,
we note that $J$ also gets shifted by terms involving $C^A$.
This shift ensures that extrema of ${\cal E}_5$ satisfy all
the five-dimensional equations of motion.  This we now demonstrate by
explicitly checking the equation of motion for $A_{5 \psi}^A$, as follows.
Using \eqref{gaugepot5}, we compute 
\begin{eqnarray}
{\cal F}_5 &=& \pi \, 
\frac{v_1 \, (v_2^{3} \, v_3)^{1/2}}{4 G_5}  
\left[
 - \frac{1}{v_1} + \frac{4 - v_3}{v_2} + \frac{v_2 \, v_3 \, \alpha^2}{16 
v_1^2} + \frac{G_{AB} \, e^A_5 \, e^B_5 }{2 v_1^2} - 
 8 R^2 \, 
\frac{G_{AB} \, C^A \, C^B}{v_2^2}  \right] \nonumber\\
&& - \frac{2 \pi}{3 G_5} \, R^2 \, C_{ABC} \, C^A \, C^B \, e^C_5
\;.
\label{f5struc}
\end{eqnarray}
Then, varying the entropy function ${\cal E}_5$ with respect
to the electric fields $e^A_5$ and setting 
$\partial_e {\cal E}_5 =0$ yields
\begin{eqnarray}
 \frac{\pi}{4 G_5}  \, \frac{(v_2^3 \, v_3)^{1/2}}{v_1} \, G_{AB} \, e^B_5
= - \frac12 \, {\hat q}_A 
\;, 
\label{varef}
\end{eqnarray}
while varying with respect to $C^A$ and setting
$\partial_C {\cal E}_5 =0$ gives
\begin{eqnarray}
%&& 
- \frac{\alpha}{2} \, {\hat q}_A 
+ \frac{2 \pi  R}{G_5} \, C_{ABC} \, C^B \, e^C_5 
%\nonumber\\ && \qquad \qquad 
+  \frac{4 \pi R}{G_5} \, \frac{v_1 (v_2^3 \,v_3)^{1/2}}{v_2^2} 
\, G_{AB} C^B =0
\;,
\label{varec}
\end{eqnarray}
where we introduced 
\begin{equation}
{\hat q}_A = q_A \, G_5^{-1/3} - \frac{2\pi R^2}{G_5} C_{ABC} \,C^B \,C^C \;, 
\end{equation}
for convenience.  
Combining \eqref{varef} and \eqref{varec} results in
\begin{eqnarray}
 \alpha  \, \frac{(v_2^3 \, v_3)^{1/2}}{v_1} \, G_{AB} \, e^B_5
+ 8 R \, C_{ABC} \, C^B \, e^C_5 + 16 R \, 
\frac{v_1 (v_2^3 \,v_3)^{1/2}}{v_2^2} \, G_{AB} C^B =0 \;,
\label{eompsi}
\end{eqnarray}
which is precisely the equation of motion for $A_{5 \psi}^A$ evaluated
in the black hole background.  Observe that when $\alpha \, q_A \neq 0$, then
generically also $C^A \neq 0$.  We also note 
that when expressed in terms of four-dimensional
quantities, ${\hat q}_A$ equals ${\hat q}_A = G_4^{-1/2} \left(Q_A - 
{\rm Re} {\cal N}_{A0} \, P^0 \right)$, where $P^0$ is given by 
\eqref{valuep0}.

The entropy function \eqref{entropyf5} depends on a set of constant parameters,
namely $e^A_5, X^A, C^A$, $v_1, v_2, v_3$ and $\alpha$, whose
horizon values are determined by extremizing ${\cal E}_5$.
To this end, we compute the (remaining) extremization equations.  
Inserting \eqref{varef} into \eqref{entropyf5}
gives
\begin{eqnarray}
{\cal E}_5  &=&  2 \pi \, 
\alpha \left[
- \frac12 J
- \frac12 R \,C^A \left( 
q_A \, G_5^{-1/3} - 
\frac{2 \pi  R^2}{ 3 G_5} \, C_{ABC} \, C^B \, C^C  \right)\right] 
+
G_5 \, \frac{v_1}{(v_2^3 \, 
v_3)^{1/2}} 
\, {\hat q}_A \, G^{AB} \,
{\hat q}_B 
\nonumber\\
&& 
%\qquad 
- \frac{\pi^2}{2 G_5} v_1 (v_2^3 \, v_3)^{1/2} \left[ - \frac{1}{v_1}
+ \frac{4 - v_3}{v_2} + \frac{v_2 \, v_3 \alpha^2}{16 v_1^2} - 8 R^2
\frac{G_{AB} C^A C^B}{v_2^2} \right] \;.
\label{entrof5}
\end{eqnarray}
Demanding $\partial_{\alpha} {\cal E}_5 =0$ results in the expression
for the angular momentum, 
\begin{equation}
\frac{\pi}{32 G_5} \frac{v_2^{5/2} v_3^{3/2}}{v_1} \, \alpha = - \frac12 J
- \frac12 R \,C^A \left( 
q_A \, G_5^{-1/3} - 
\frac{2 \pi  R^2}{ 3 G_5} \, C_{ABC} \, C^B \, C^C  \right)
\;.
\label{valuealp}
\end{equation}
Computing $\partial_{v_i} {\cal E}_5 = 0$ (with $i =
1,2,3$), we obtain
\begin{eqnarray}
v_1 &=& \frac{v_2}{4} \;, \nonumber\\
v_2 \, v_3 \left[ 2 v_2 + v_2 \, v_3 (1 - 2 \alpha^2) \right] &=&
\frac{2 G_5^2}{\pi^2} 
%\left[G_5 
\, {\hat q}_A \, G^{AB} \, {\hat q}_B 
\;,
\nonumber\\
2 v_2 - v_2 v_3 (2 - \alpha^2) &=& 8 R^2 \,G_{AB} \, C^A \, C^B \;.
\label{vvs}
\end{eqnarray}
Observe that the first of these conditions yields ${\tilde v}_1 = {\tilde
v}_2$, as can be seen from \eqref{v1v2}.  This
implies the vanishing of the Ricci scalar for the  associated four-dimensional 
geometry.

Inserting the relations \eqref{valuealp} and \eqref{vvs} into \eqref{entrof5} 
results in
\begin{eqnarray}
{\cal E}_5 = \frac{\pi^2}{2 G_5} \left(v_2^3 \, v_3 \right)^{1/2} \;,
\label{extrent}
\end{eqnarray}
which exactly equals the macroscopic entropy ${\cal S}_{\rm macro} = 
A_5/(4 G_5)$ of the rotating black hole, where $A_5$ denotes the horizon
area.

Introducing the abbreviations
\begin{eqnarray}
\Omega &=& 
\frac{2 G_5^2}{\pi^2}\, \frac{1}{\sqrt{v_2 \, v_3}} \, 
% \left[G_5 \, 
{\hat q}_A \, G^{AB} \, {\hat q}_B 
\;,
\nonumber\\
\Delta &=& 8 R^2 \, \sqrt{v_2 \, v_3} \, 
G_{AB} \, C^A \, C^B \;, \nonumber\\
\Gamma &=& \frac{8 G_5}{\pi} \, \left[
- \frac12 J
- \frac12 R \,C^A \left( 
q_A \, G_5^{-1/3} - 
\frac{2 \pi  R^2}{ 3 G_5} \, C_{ABC} \, C^B \, C^C  \right) \right]
\;,
\label{ogd}
\end{eqnarray}
we obtain from \eqref{valuealp} and \eqref{vvs} the following
two equations,
\begin{eqnarray}
3 (v_2 \, v_3)^{3/2}
- 3 \frac{\Gamma^2}{(v_2 v_3)^{3/2}}
&=&  \Omega - \Delta \;, \nonumber\\
\sqrt{v_2 \, v_3} \left( 6 v_2 - 3 v_2 v_3 \right) &=& \Omega + 2 \Delta \;.
\label{v2v3odg}
\end{eqnarray}
Solving the first of these equations yields (with $v_2 v_3$ positive)
\begin{eqnarray}
(v_2 \, v_3)^{3/2} = \frac16 \, (\Omega - \Delta) +  \sqrt{\Gamma^2
+ \frac{1}{36} (\Omega - \Delta)^2} \;.
\label{v2v332}
\end{eqnarray}
Inserting this into the second equation of \eqref{v2v3odg} gives
\begin{equation}
\left(v_2^3 \, v_3 \right)^{1/2} = \frac14 \left(\Omega + \Delta \right)
+ \frac12 \sqrt{\Gamma^2
+ \frac{1}{36} (\Omega - \Delta)^2} \;.
\label{area5}
\end{equation}
Thus, by taking suitable ratios of \eqref{v2v332} and \eqref{area5},
we obtain $v_2$ and $v_3$ expressed in terms of $\Omega, \Delta$ and
$\Gamma$.  Now, recalling the definition of ${\hat X}^A$ in \eqref{dic}
and using \eqref{vpvv},
we have $\sqrt{v_2 \, v_3} \, G_{AB} = 2 R \, {\hat G}_{AB}$, where
\begin{equation}
{\hat G}_{AB} = -  C_{ABC} \, {\hat X}^C + 9 \,\frac{{\hat X}_A \,
{\hat X}_B}{{\hat V}}
\;,
\label{ghatxh}
\end{equation}
with ${\hat X}_A$ and ${\hat V}$ defined in \eqref{hatxphi}.
Therefore $\Omega, \Delta$ and $\Gamma$, 
and hence also the horizon area \eqref{area5},
 are entirely
determined in terms of the scalar fields ${\hat X}^A$ and $C^A$
and the charges.  The horizon values of ${\hat X}^A$ and $C^A$
are in turn determined in terms of the charges by 
solving the respective extremization
equations.  The extremization equations for the $C^A$ are 
given by \eqref{varec}, while the extremization equations for the
${\hat X}^A$ are obtained by setting $\partial_{{\hat X}^A} {\cal E}_5 = 0$.
Rather than computing the horizon values in this way, we will
determine them
by solving the associated attractor equations in four dimensions.
This will be done in the next subsection.  

Finally, let us consider static black holes.  
When the rotation parameter
$\alpha$ is set to zero, we have 
$\Gamma = 0$ and \eqref{eompsi} can be abbreviated as
$D_{AB} \, C^B = 0$.  In the following we will 
assume that $D_{AB}$ is invertible
so that
$C^A = 0$.  We then infer from \eqref{vvs} and \eqref{ogd} that
$v_3=1$, 
$\Delta = 0$ and 
\begin{equation}
\Omega = 
\frac{ G_5^{4/3}}{\pi^2 R}\, 
{q}_A \, {\hat G}^{AB} \, {q}_B \;,
\end{equation}
which is the black hole potential in five dimensions for 
static electrically charged black holes 
\cite{Ferrara:2006xx}.
Using \eqref{char45}, we obtain for \eqref{extrent},
\begin{equation}
{\cal E}_5 = \frac{2 \pi}{3} \, Q_A \, {\hat G}^{AB} \, Q_B \;.
\label{5dbhp}
\end{equation}
{From} \eqref{coup45} and \eqref{ghatxh} 
we infer that ${\hat G}_{AB} = - {\rm Im} {\cal N}_{AB}$.
With the help of \eqref{coup45}, 
\eqref{hatxphi} and \eqref{area5} we compute
\begin{equation}
{\rm Im} {\cal N}_{00} = -
{\hat V} =  \frac{1}{12 \pi } \frac{G_5}{R^3 } \, 
Q_A \, [({\rm Im} {\cal N})^{-1}]^{AB} \, Q_B \;,
\end{equation}
where we used ${\rm Im} {\cal N}_{A0} = 0$.  It follows that we can rewrite
\eqref{5dbhp} as
\begin{equation}
{\cal E}_5 = -\frac{2 \pi}{4} \left[
(P^0)^2 \, {\rm Im} {\cal N}_{00} + Q_A \, 
[({\rm Im} {\cal N})^{-1}]^{AB} \, Q_B \right] \;,
\end{equation}
where
\begin{equation}
P^0 = p^0 \, \frac{R}{G_4^{1/2}} \;\;\;,\;\;\; p^0 = 1 \;.
\label{valuep0}
\end{equation}
Thus, \eqref{5dbhp} precisely equals the four-dimensional black hole
potential \cite{Ferrara:1997tw,Gibbons:1997cc,Goldstein:2005hq},
\begin{equation}
{\cal E}_4 = -\frac{2 \pi}{4} 
(Q_I - {\cal N}_{IK}  P^K ) \, [({\rm Im} {\cal N})^{-1}]^{IJ} \, 
(Q_J - {\bar {\cal N}}_{JL} P^L ) \;,
\label{fourentf}
\end{equation}
for the case at hand with $C^A = 0$ and non-vanishing
charges $(P^0, Q_A)$, as it should.  In \eqref{fourentf} 
$(P^I, Q_J)$ denote the magnetic and electric charges in four
dimensions, respectively.

\subsection{Attractor equations and examples}

The four-dimensional entropy function \eqref{fourentf} can be 
rewritten into \cite{Cardoso:2006xz}
\begin{eqnarray}
{\cal E}_4  = \pi \left[ \Sigma + \left( {\cal Q}_I - F_{IJ} {\cal P}^J \right)
N^{IK}  \left( {\cal Q}_K - {\bar F}_{KL} {\cal P}^L \right) \right]\;,
\label{entrof4d}
\end{eqnarray}
where
\begin{eqnarray}
\Sigma &=& - \im \left( {\bar Y}^I F_I - Y^I {\bar F}_I \right) 
- Q_I (Y^I + {\bar Y}^I ) + P^I (F_I + {\bar F}_I ) 
\;, \nonumber\\
N_{IJ} &=& \im \left( {\bar F}_{IJ} - F_{IJ} \right) \;, \nonumber\\
{\cal Q}_I &=& Q_I + \im (F_I - {\bar F}_I)  \;, \nonumber\\
{\cal P}^I &=& P^I + \im (Y^I - {\bar Y}^I) \;.
\end{eqnarray}
For the notation cf. appendix \ref{appred}.
The scalar fields \eqref{zA} are expressed in terms of the $Y^I$
by $z^A = Y^A / Y^0$.
The horizon values of the scalar fields ${\hat X}^A$ and $C^A$ can
be conveniently determined by solving the attractor equations for the $Y^I$
in four dimensions, which read \cite{Cardoso:2006cb,Cardoso:2006xz}
\begin{eqnarray}
- 2 ( {\cal Q}_J - F_{JK}\, {\cal P}^K ) 
+ \im  ( {\cal Q}_I - {\bar F}_{IM}\, {\cal P}^M ) N^{IR} \, F_{RSJ} \, N^{SK}
( {\cal Q}_K - {\bar F}_{KL}\, {\cal P}^L )  = 0 \;.
\label{attractor4d}
\end{eqnarray}
Contracting with $Y^I$ results in 
\begin{eqnarray}
\im \left({\bar Y}^I \, F_I 
- Y^I \, {\bar F}_I \right) = P^I \, F_I - Q_I \, Y^I \;.
\label{conatt}
\end{eqnarray}
Supersymmetric black holes satisfy ${\cal Q}_I = {\cal P}^J =0$.

In the following, we will discuss two classes of four-dimensional
non-supersymmetric
extremal black holes which are connected to five-dimensional black holes.
These have a non-vanishing $P^0$ given by \eqref{valuep0}.
The first class consists of black holes with non-vanishing
charges $(P^0, Q_A)$ in heterotic-like theories with prepotential
$F(Y) = - Y^1 Y^a \eta_{ab} Y^b/Y^0$, 
where $\eta_{ab}$ denotes a symmetric matrix with the
inverse $\eta^{ab}$ ($\eta^{ab} \eta_{bc} = \delta^a_c$)
and $a,b = 2, \dots, n$.  These black holes are static 
in five dimensions.  Taking $P^0 >0$ and $Q_1 \, Q_a \, \eta^{ab} \, Q_b < 0$, 
we find that the 
attractor equations \eqref{attractor4d} are solved by 
\begin{eqnarray}
Y^0 &=& - \frac{\im}{4} \, P^0 \;, \nonumber\\
Y^1 &=&  \frac18 \, \sqrt{-\frac{P^0 \, Q_a \, \eta^{ab} \, Q_b}{ Q_1}} \;,
\nonumber\\
Y^a &=& - \frac14 \,\sqrt{- \frac{P^0 \, Q_1 }{Q_c \, \eta^{cd} \, Q_d}} 
\; \eta^{ab} \, Q_b\;.
\label{hetys}
\end{eqnarray}
The $z^A$ read, 
\begin{eqnarray}
z^1 &=& \im \, {\hat X}^1 = 
\frac{\im}{2} \sqrt{- \frac{Q_a \, \eta^{ab} \, Q_b}{P^0 \, Q_1}}
\;, \nonumber\\
z^a &=&  \im \, {\hat X}^a = 
- \im \,\sqrt{- \frac{Q_1 }{P^0 \, Q_c \, \eta^{cd} \, Q_d}} 
\; \eta^{ab} \, Q_b  \;.
\end{eqnarray}
Requiring ${\hat V} > 0$ for consistency 
(see \eqref{hatxphi}) restricts the charges to 
$Q_a \, \eta^{ab} \, Q_b > 0$ and $Q_1 < 0$.
Using \eqref{hetys}, \eqref{entrof4d}, 
\eqref{valuep0} and \eqref{char45}, the entropy is computed to be
\begin{eqnarray}
{\cal E}_5 = \pi \sqrt{- P^0 \, Q_1 \, Q_a \, \eta^{ab} \, Q_b} 
= \frac{\sqrt{\pi}}{2} \, \sqrt{ - q_1 \, q_a \, \eta^{ab} \, q_b} 
\;.
\label{entrohet45}
\end{eqnarray}
Upon performing the rescaling $q_A \rightarrow (4 \pi)^{1/3} \, q_A$,
the entropy \eqref{entrohet45} attains its standard form.
For the case $n=3$ with 
non-vanishing $\eta_{23}= \eta_{32}=\frac12$, the so-called STU model,
the above solution
has been given in \cite{Kallosh:2006ib} and found to be stable.  Requiring
the moduli $S,T$ and $U$ to lie in the K\"ahler cone imposes the
additional restriction $Q_2 < 0$ and $Q_3 < 0$.

The solution \eqref{hetys} is non-supersymmetric in four dimensions,
since ${\cal Q}_A \neq 0, {\cal P}^0 \neq 0$.  We now check 
the supersymmetry of the associated five-dimensional solution.
An electrically charged 
supersymmetric solution in five dimensions satisfies the condition
${\cal A}_A = 0$ 
\cite{Chamseddine:1996pi,Chou:1997ba,Sabra:1997yd,Chamseddine:1998yv},
where in our conventions (see appendix \ref{appred})
\begin{equation}
{\cal A}_A = q_A - 2 \, {\rm e}^{6 \phi} \, Z({\hat X})  \, {\hat X}_A  
\;\;\;,\;\;\;Z ({\hat X})  = q_A \, {\hat X}^A \;,
\label{calaa}
\end{equation}
with ${\hat X}_A$ given in \eqref{hatxphi}.
Computing ${\cal A}_A$ for the solution \eqref{hetys} using \eqref{char45}, 
we find that
${\cal A}_A = G_5^{1/3} \, G_4^{-1/2} \, (Q_A + 3 P^0 \, {\hat X}_A) =0$.
The entropy \eqref{entrohet45} takes the supersymmetric form 
${\cal E}_5= (2 \pi)^{1/2} \, 3^{-3/2} \, |q_A \, X^A|^{3/2}$.
Solutions which are supersymmetric
from a higher-dimensional point of view, but non-supersymmetric from
a lower-dimensional point of view, have been
discussed in \cite{Nilsson:1984bj,Duff:1997qz}
and occur when dimensionally reducing geometries that
are $U(1)$-fibrations, as in our case.  In string theory one can
generate new solutions from a given one
by using duality transformations.
Two configurations which 
are related in this manner must be both
supersymmetric or both non-supersymmetric in four dimensions.
Hence, the configurations obtained in this way from the $(P^0, Q_A)$ solution 
\eqref{hetys} will be non-supersymmetric
in four dimensions.
Those with a positive $P^0$
can be lifted to five-dimensional solutions which, 
depending on the specific duality transformation, 
may or may not be supersymmetric.

The second class of solutions we consider
consists of black holes with non-vanishing charges
$(P^0, Q_0)$.  They correspond to rotating black holes in five dimensions, of
the type discussed in \cite{Rasheed:1995zv,Itzhaki:1998ka,Larsen:1999pp,
Astefanesei:2006dd,
Emparan:2006it,Astefanesei:2006sy,Dabholkar:2006tb}, which are not
supersymmetric.
We use the 
prepotential $F(Y) = - Y^1 \, Y^2 \, Y^3 /Y^0$.
Taking $P^0 \,Q_0>0$ and ${\rm Re} z^A = 0$, we find
that the attractor equations \eqref{attractor4d} are solved by 
\begin{eqnarray}
Y^0 &=& - \frac{( 1 - \im)}{8} \, P^0 \;, \nonumber\\
Y^1 \, Y^2 \, Y^3 &=& \im (1 - \im)^3 \, \frac{(P^0)^2 \, Q_0 }{512} \;.
\label{solpq0}
\end{eqnarray} 
Observe that the attractor equations do not determine the individual values
$Y^1, Y^2$ and $Y^3$, because the entropy function has two flat directions.
The coupling constant
 ${\rm Im} {\cal N}_{00}$
and the entropy are,
however, determined in terms of the charges.  The former takes the value
$-{\rm Im} {\cal N}_{00} = \im z^1 \, z^2 \, z^3 =  Q_0/P^0 > 0$.
We also find that ${\hat V} > 0$, as required by consistency.
{From} \eqref{v2v332} and \eqref{area5} 
we obtain $v_2 = \frac12 \, |\Gamma|^{2/3}$
and $v_3 = 2$, and from 
\eqref{valuealp} we have $\alpha = {\rm sgn} \,\Gamma$.
Using \eqref{solpq0}, \eqref{entrof4d}, 
\eqref{valuep0} and \eqref{char45}, the entropy is computed to be
\begin{eqnarray}
{\cal E}_5 = \pi \, P^0 \, Q_0 = \pi \, J \;.
\end{eqnarray}

We close this section by displaying the relation 
between the five-dimensional quantity $Z( {\hat X})  \, {\rm e}^{6 \phi}$
appearing in \eqref{calaa} and the four-dimensional
$Y^0$ for the case of static 
black holes with $C^A =0$.
{From}  \eqref{conatt} we obtain (with $Q_0 = P^A=0$, and with $P^0$
given  by \eqref{valuep0}) 
\begin{eqnarray}
Z( {\hat X})  \, {\rm e}^{6 \phi} 
= \frac{\im}{2} \, \frac{G_5^{1/3}}{G_4^{1/2}}
\left( - 8 \, Y^0 + \im \frac{R}{\sqrt{G_4}} \right) \;,
\label{z4p}
\end{eqnarray}
where we used $Y^0 = - {\bar Y}^0$, which follows from the reality
of \eqref{z4p}.  For a supersymmetric solution in four dimensions,
${\cal P}^0 =0$ and hence $Y^0 = \im P^0/2$, so that
\begin{eqnarray}
Z( {\hat X})  \, {\rm e}^{6 \phi} 
= 6 \pi \, R^2 \, G_5^{-2/3} \;.
\label{z4psus4}
\end{eqnarray}

\section{Conclusions}

In the context of $N=2$ supergravity theories with cubic prepotentials,
we used 
the relation between extremal black hole solutions in four and in
five dimensions to define the entropy function for rotating 
extremal black holes 
in terms of the entropy function for static black holes in four dimensions.
We focused on rotating electrically charged black holes with one
independent angular momentum parameter, for simplicity, and
we discussed two classes of solutions.
General charged static black holes in four dimensions
also carry magnetic charges $P^A$, and these charges can be easily incorporated
into the discussion given above by adding a term $- P^A \, R \, \cos \theta
\, d \varphi$ to both \eqref{gaugepot5} and \eqref{gaugepot4}.
Their entropy function
is given by \eqref{entrof4d}, and the
entropy function of the associated
five-dimensional rotating black holes is then defined by \eqref{e5=e4}.

In five dimensions, rotating extremal black holes may carry two independent
angular momentum parameters \cite{Myers:1986un}.  
These black holes will be connected to rotating
extremal black holes in 
four dimensions.  The entropy function of these five-dimensional
black holes can then again be defined in terms of the entropy function
of the associated rotating four-dimensional black holes.
The entropy function for rotating attractors in four dimensions has recently
been discussed in \cite{Astefanesei:2006dd}.

%%%%%%%%%%
%\newpage
%%%%%%%%%%
%%%%%%%%%%%%%%%%%%%%%%%%%%%%%%%%%%%%%%%%%%%%%%%%%%%%%%%%%%%%%%%%%
\subsection*{Acknowledgements}
We thank G. Curio, B. de Wit, J. Rosseel, A. Van Proeyen and M. Zagermann
for useful discussions.  
This work is partly supported by EU
contract MRTN-CT-2004-005104.
%%%%%%%%%%%%%%%%%%%%%%%%%%%%%%%%%%%%%%%%%%%%%%%%%%%%%%%%%%%%%%%%%

\appendix
\section{$N=2$ supergravity actions and dimensional reduction
\label{appred}}

Here we review various elements of $N=2$ supergravity theories 
in four and in five dimensions.  
We also review the reduction of the five-dimensional action based on
very special geometry to the four-dimensional action based on special
geometry.  This will explain our conventions, 
which differ slightly from the ones used 
in \cite{Gunaydin:1983bi,deWit:1992cr,Gunaydin:2005df,Behrndt:2005he}.  
For notational simplicity, we drop
the subscripts on the five- and four-dimensional gauge fields.

The five-dimensional $N=2$ supergravity action is based on the
cubic polynomial \cite{Gunaydin:1983bi}
\begin{eqnarray}
V = \frac16 C_{ABC}  X^A X^B X^C \;,
\label{constrV}
\end{eqnarray}
where the $X^A$ are real scalar fields satisfying the constraint 
$V = {\rm constant}$.  
The five-dimensional gauge couplings $G_{AB} (X)$ are given by
\begin{eqnarray}
G_{AB} (X) = - \frac12 \partial_A \partial_B \log V 
\vert_{V = {\rm constant}} \;,
\end{eqnarray}
and hence,
\begin{eqnarray}
G_{AB} (X) = V^{-1} \left( - \frac12 C_{ABC} X^C + \frac92 \frac{X_A X_B}{V}
\right) \;,
\label{gabmetric}
\end{eqnarray}
where we defined
\begin{eqnarray}
 X_A = \frac16 C_{ABC}  X^B X^C \;.
\end{eqnarray}
Observe that
\begin{eqnarray}
G_{AB} \, X^A  X^B = \frac32 \;\;\;,\;\;\; X^A \partial_i X_A = 0 \;.
\end{eqnarray}
Here $\partial_i X^A = \frac{\partial}{\partial \varphi^i} X^A (\varphi)$,
where $\varphi^i$ denote the physical scalar fields with target space
metric $g_{ij} = G_{AB} \, \partial_i X^A \, 
\partial_j X^B$.

The bosonic part of the five-dimensional $N=2$ supergravity action reads
\begin{eqnarray}
S_5 &=& \frac{1}{8 \pi G_5} \, \left[ \int d^5 x 
\sqrt{-G } \left( \frac12 R_G 
- \frac12 \,
G_{AB} \, \partial_M X^A \,\partial^M X^B
%g_{ij} \, \partial_M \phi^i \,\partial^M \phi^j 
- \frac14 G_{AB} \, F^A_{MN} F^{B MN} \right) \right. \nonumber\\
&& \left. \qquad \qquad - \frac{1}{6} \int C_{ABC} \, F^A \wedge F^B \wedge A^C
\right]
\;,
\label{lag5}
\end{eqnarray}
where $G$ denotes the determinant of the spacetime metric in five dimensions.

The four-dimensional 
$N=2$ supergravity action, on the other hand, is based on the
prepotential \cite{deWit:1984pk,deWit:1984px}
\begin{eqnarray}
F(Y) = - \frac16 \, \frac{C_{ABC} \, Y^A Y^B Y^C}{Y^0} \;,
\label{prepotential}
\end{eqnarray}
where the $Y^I$ are complex scalar fields ($I=0, A$).
The four-dimensional gauge couplings ${\cal N}_{IJ}$ are given by
\begin{eqnarray}
\mathcal{N}_{IJ}
=\bar F_{IJ}+2 \im \frac{\Im F_{IK} \, \Im F_{JL} \, Y^K Y^L}{\Im F_{MN} 
\, Y^M 
Y^N}\;,
\label{nij}
\end{eqnarray}
where $F_I = \partial F/\partial Y^I \,,\, F_{IJ} = \partial^2 F/\partial Y^I
\partial Y^J$.  The four-dimensional physical scalar fields $z^A$ 
are 
\begin{equation}
z^A = \frac{Y^A}{Y^0} \;.
\end{equation}
The bosonic part of the four-dimensional $N=2$ supergravity action reads
\begin{eqnarray}
S_4 &=& \frac{1}{8 \pi G_4} \int d^4 x \,  \sqrt{-g} \, \left( \frac12 R_g - 
g_{A \bar B} \, \partial_{\mu} z^A \partial^{\mu} 
{\bar z}^B + \frac14 \, {\rm Im} {\cal N}_{IJ} \, F_{\mu \nu}^I \,
F^{J \mu \nu} \right. \nonumber\\
&& \left.  \qquad \qquad \qquad \qquad \qquad \qquad \qquad \qquad \qquad \;
- \frac14 
{\rm Re} {\cal N}_{IJ} \, F_{\mu \nu}^I \,
{\tilde F}^{J \mu \nu} 
\right) \;,
\label{lag4}
\end{eqnarray}
where
$g$ denotes the determinant of the spacetime metric in four dimensions,
and where 
${\tilde F}^{J ab} = \frac12 \varepsilon_{abcd} F^{J cd}$ with 
$\varepsilon_{0123} = 1$.
The quantity $g_{A \bar B}$ is the K\"ahler metric $g_{A \bar B}
= \frac{\partial}{\partial z^A} \frac{\partial}{\partial {\bar z}^B} K$
computed from the K\"ahler potential $K(z, {\bar z})$.
For the prepotential \eqref{prepotential}, the K\"ahler potential
reads
\begin{equation}
{\rm e}^{-K (z, {\bar z})} = \frac{\im }{6} C_{ABC} (z^A - {\bar z}^A)
 (z^B - {\bar z}^B) (z^C - {\bar z}^C) \;.
\label{kahlerpot}
 \end{equation}

Now we perform the reduction of \eqref{lag5} along
$x^5$ down to four dimensions using \eqref{dic}.
We take the various fields to be independent of the fifth coordinate
$x^5$.
Setting $x^5 = R \psi \,,\, 0 \leq \psi < 4 \pi$, we use that
the five- and four-dimensional Newton constants are related by 
\begin{equation}
G_5 = 4 \pi \, R \, G_4 \;.
\label{gn45}
\end{equation}
Reducing the gauge kinetic terms $G_{AB} F^A F^B$ gives rise to 
a scalar kinetic term of the form
\begin{equation}
- \frac14 \sqrt{-G} \, G_{AB} \, F^A F^B  \rightarrow -\frac12 \, \sqrt{-g} \, 
\ee^{4 \phi} \,G_{AB} \,
\partial_{\mu} C^A \partial^{\mu} C^B \;,
\label{cc}
\end{equation}
whereas reducing $R_G - G_{AB} \partial_M X^A \partial^M X^B$ gives rise
to scalar kinetic terms for ${\hat X}^A = {\rm e}^{- 2\phi} \, X^A$,
\begin{eqnarray}
\sqrt{-G } \left( \frac12 R_G 
- \frac12 G_{AB} \partial_M X^A \partial^M X^B 
\right) \rightarrow
\sqrt{-g } \left( \frac12 R_g 
- \frac12 
{\rm e}^{4 \phi} \,G_{AB} \, 
\partial_{\mu} {\hat X}^A \,\partial^{\mu} {\hat X}^B
\right) \;.
\label{hatxhatx}
\end{eqnarray}
Eqs. \eqref{cc} and \eqref{hatxhatx} can be combined into
\begin{eqnarray}
\sqrt{-g } \left( \frac12 R_g 
- \frac12 \,
{\rm e}^{4 \phi} \,G_{AB} \,
 \partial_{\mu} z^A \, \partial^{\mu} {\bar z}^{\bar B}
\right) \;,
\label{sca45}
\end{eqnarray}
where $z^A$ is defined as in \eqref{zA}.  Using \eqref{kahlerpot}
we compute $g_{A {\bar B}} = \frac12 {\rm e}^{4 \phi} \, G_{AB}$,  
and hence \eqref{sca45} can be written as
\begin{eqnarray}
\sqrt{-g } \left( \frac12 R_g 
- g_{A {\bar B}} \,
 \partial_{\mu} z^A \, \partial^{\mu} {\bar z}^{\bar B}
\right) \;.
\end{eqnarray}
In addition, 
reducing $R_G$ and $G_{AB} F^A F^B$ also gives rise to 
the four-dimensional gauge kinetic terms
\begin{eqnarray}
\label{f4f5lag}
\sqrt{-G} \, R_G &\rightarrow& 
\sqrt{-g } \left( R_g - \frac14 {\rm e}^{- 6 \phi} \,
F^0 \, F^0 \right) \;, \\
- \frac12 \,\sqrt{-G} \, G_{AB} \, F^A \, F^{B}  &\rightarrow&
- \frac12 \,
\sqrt{-g} \, {\rm e}^{-2 \phi} \, G_{AB}
\left[ F^A F^B - 2 C^B \, F^A \, F^0 + C^A \, C^B \, F^0 F^0 
\right] \nonumber \;.
\end{eqnarray}
This we compare with 
${\rm Im} {\cal N}_{IJ} \, F^I \, F^J$ in four dimensions.  
To this end, we
compute the couplings ${\cal N}_{IJ}$ for the prepotential 
\eqref{prepotential} and we express them in terms of the fields ${\hat X}^A$
and $C^A$ using \eqref{zA},
\begin{eqnarray}
 {\cal N}_{00} &=& - \frac{1}{3 } \, C_{ABC} \, C^A C^B C^C -
\im \left[
2 \,{\rm e}^{- 2 \phi} \, V\, G_{AB} \, C^A C^B + 
{\hat V} \right] \;, \nonumber\\
 {\cal N}_{0A} &=& \frac{1}{2 } \,C_{ABC} \, C^B C^C + 2\im 
\,{\rm e}^{- 2 \phi} \, V\, G_{AB} \,
C^B \;,  \nonumber\\
{\cal N}_{AB}  &=& -  \, C_{ABC} \, C^C -
2\im \,{\rm e}^{- 2 \phi} \, V\, G_{AB} \;,
\label{coup45}
\end{eqnarray}
where
\begin{eqnarray}
{\hat {V}} = {\hat X}_A \, {\hat X}^A \;\;\;,\;\;\; {\hat X}_A = 
\frac16 C_{ABC} {\hat X^B }{\hat X^C} \;\;\;,\;\;\;
{\rm e}^{-6 \phi} = V^{-1} \, {\hat V} \;. 
\label{hatxphi}
\end{eqnarray}
Hence we find that the sum of the field strength
terms on the right hand side of \eqref{f4f5lag} equals
\begin{equation}
 \frac{1}{4V} \, {\rm Im} {\cal N}_{IJ} \, F^I \, F^J \;.
\end{equation}
Thus, requiring the matching of the five-dimensional gauge
kinetic term $- \frac14 G_{AB} F^A F^B$ in \eqref{lag5}
with the four-dimensional
gauge kinetic term $\frac14 {\rm Im} {\cal N}_{IJ} \, F^I \, F^J $
in \eqref{lag4}
yields the normalization condition
\begin{equation}
2 \, V = 1 \;.
\label{valuev}
\end{equation}
Next, we reduce the five-dimensional Chern-Simons term 
$C_{ABC} \, F^A \wedge  F^B \wedge A^C $
in \eqref{lag5}.  Using \eqref{dic}, we first observe
that $C_{ABC} \, F^A \wedge  F^B \wedge A^C_{\psi} d \psi  $
can be expressed in terms of four-dimensional gauge fields as,
\begin{eqnarray}
C_{ABC} \, F^A \wedge  F^B \wedge A^C_{\psi} \,d \psi 
&=& R \, C_{ABC} \left[ \right. C^A \, F^B  \wedge F^C  
- C^A \, C^B \, F^C  \wedge F^0 \nonumber\\
&& \qquad \qquad + \frac13 C^A \, C^B \, C^C \, F^0 
\wedge F^0 \left. \right] \wedge d \psi \;, 
\label{cspsi}
\end{eqnarray}
up to a total derivative term.  The field strengths 
on the right hand side are four-dimensional, and $ A^C_{\psi} = R \,
C^C$.  Using 
\begin{equation}
C_{ABC} \, 
C^A \, F^B  \wedge F^C \wedge d\psi  = - \frac12 
 d\psi \, d^4x \,
\sqrt{-g} \, C_{ABC} \, C^A \, F^B 
\, {\tilde F}^C \;,
\end{equation}
and similarly for the other terms in \eqref{cspsi}, we obtain
(up to a total derivative)
\begin{eqnarray}
C_{ABC} \, F^A \wedge  F^B \wedge A^C_{\psi} \,d \psi 
= \frac12 \, R \, d\psi \, d^4x \, \sqrt{-g} 
\, {\rm Re} {\cal N}_{IJ} \, F^I \, {\tilde F}^J
\;,
\end{eqnarray}
where we used \eqref{coup45}.  Then, using
\begin{equation}
C_{ABC} \, F^A \wedge  F^B \wedge A^C = 3 
C_{ABC} \, F^A \wedge  F^B \wedge A^C_{\psi} \,d \psi \;,
\end{equation}
which holds up to a total derivative term, we obtain
\begin{equation}
\frac{1}{6 G_5} \int C_{ABC} \, F^A \wedge  F^B \wedge A^C = 
\frac{1}{4 G_4}   \, \int \, d^4x \, \sqrt{-g} 
\, {\rm Re} {\cal N}_{IJ} \, F^I \, {\tilde F}^J
\;.
\end{equation}
Thus, dimensional reduction of \eqref{lag5} yields \eqref{lag4}, up to
boundary terms.

%%%%%%%%%%%%%%%%%%%%%%%%%%%%%%%%%%%%%%%%%%%%%%%%%%%%%%%%%%%%%%%%%%%%
\providecommand{\href}[2]{#2}
\begingroup\raggedright\endgroup

\end{document}